\documentstyle[preprint,prb,aps]{revtex}

\begin{document}
\title{Superconductivity in sintered-polycristalline PrBa$_{2}$Cu$_{3}$O$_{7-\delta
}$}
\author{F. M. Araujo-Moreira$^{a,*}$P. N. Lisboa Filho$^{a}$, S. M. Zanetti$^{b}$,
E. R. Leite$^{b}$, and W. A. Ortiz$^{a}$}
\address{$^{a}$ Grupo de Supercondutividade e Magnetismo, Departamento de F\'{i}sica\\
$^{b}$ Lab. Interdisciplinar de Eletroqu\'{i}mica e Cer\^{a}mica,
Departamento de Qu\'{\i }mica\\
Universidade Federal de S\~{a}o Carlos, Caixa Postal 676, S\~{a}o Carlos-SP,
13565-905 Brazil}


\date{\today}

\maketitle

\begin {abstract}

Superconductivity in thin films and powders, and in single crystals of
 Pr-123 has been found by Blackstead et al. and Zou  et al. , respectively. 
Nevertheless, up to now it has  never been reported in sintered-policrystalline samples. 
We have prepared high-quality samples of this material by following a sol-gel method.
 We characterized  the structure of all samples by using XRD and SEM-EDS
 techniques. Magnetic characterization was performed by measuring the magnetization 
as a function of the temperature $T$, and the applied magnetic field $H$. 
Measurements were taken in the ranges 2 K <T < 400 K and  0 <H< 5 Tesla. 
In this work we report by the first  time superconductivity in Pr-123
sintered-polycristalline samples, with $T{_c}$ around 90 K, and $H{_c1}$
around 870 Oe.


\pacs{345}
\end{abstract}


Since the early years of high-temperature superconductivity, all rare earths
(RE) participating in the structure REBa$_{2}$Cu$_{3}$O$_{7-\delta }$ were
found to be superconductors, with one exception: RE=Pr, forming PrBa$_{2}$Cu$%
_{3}$O$_{7-\delta }$, also called PBCO. However, in 1995, Blackstead {\it et
al.} \cite{blackstead2} verified superconductivity in PBCO powders and in
thin films grown by pulsed laser deposition (PLD). They pointed out that
perfect PBCO should be superconductor in the vicinity of the Cu-O chains, so
that cuprate-plane models of superconductivity should be invalid. On the
other hand, in 1997 Zou {\it et al.} detected superconductivity in PBCO
single crystals \cite{zou2}, grown by both the slow-cooling and the
travelling-solvent floating-zone methods. Nevertheless, it has never been
reported superconductivity in sintered-policrystalline PBCO samples. This
has been considered consistent with previous works pointing out the high
sensitivity of the properties of PBCO to the synthesis conditions. This lack
of superconducting polycristalline PBCO samples has been one of the main
points supporting the theory that considers the suppresion of $T_{c}$ as a
consequence of the presence of Ba-site Pr.

We report in this work, by the first time, superconductivity in high-quality
sintered-polycristalline PBCO samples, obtained by a sol-gel technique.

Magnetic characterization was performed by using a Quantum Design-MPMS-5
SQUID magnetometer through $M(T)$ and $M(H)$ experiments, for different
values of the applied magnetic field and temperature. $M(T)$ were performed
in zero-field cooling (ZFC) conditions. These results were obtained (Fig. 1)
by substracting the paramagnetic background. To do that, we have considerd
that, for temperatures high enough ($T>200$ K), superconductivity has been
suppressed, remaning the Curie-Weiss paramagnetic background. For the $M(H)$
experiments (Fig.2) we have also substracted a paramagnetic background
consisting of a straight line. The line coefficients were calculated from a
pure paramagnetic behavior present in high fields ($H>4$T), and at
temperatures higher than $T_{c}$ ($T>200$ K). The results obtained through
this procedure are in perfect agreement with those expected for a
superconductor. Curves of $M(T)$ show broad superconducting transitions,
with $T_{c}$ around $90$ K for all measured samples (Fig.1). Curves of $M(H)$
show a linear diamagnetic behavior up to around 870 Oe for $T=30$ K, which
should correspond to the lower critical magnetic field, $H_{c1}$ (Fig.2).
Contributions from sample holder has already been substracted. From these
experiments we have estimated in about 5\% the superconducting fraction.

In summary, in this work we report by the first time superconductivity in
polycristalline PBCO samples, obtained by a sol-gel route. The quality of
the samples has been verified through X-ray diffraction and scanning
electron microscopy with local chemical analysis. All magnetic experiments
performed, magnetization as a function of temperature for different values
of the applied magnetic field, and magnetization as a function of the
applied magnetic field, for different values of the temperature, have
revealed the typicall behavior of a superconductor. The obtained results for
the critical temperature and lower critical magnetic field are in perfect
agreement with those reported for single crystals and thin films of PBCO.

We gratefully acknowledge brazilian agencies CNPq and FAPESP for financial
support. PNLF and SMZ thank FAPESP for grants 96/05683-1 and 96/10118-1,
respectively. FMAM gratefully acknowledges FAPESP for financial support
through grant No. 99/04393-8.

(*) To whom all correspondence should be addressed; e-mail:
faraujo@power.ufscar.br

Figure Captions

Fig.1 ZFC magnetization as a function of temperature $M(T)$, for an external
field $H=100$ Oe.

Fig. 2 Magnetization as a function of the external magnetic field $M(H)$ for
a temperature of $T=30K$.


\begin{references}
\bibitem{blackstead2}  H.A.Blackstead, D.B.Chrisey, J.D.Dow, J.S.Horwitz,
A.E. Klunzinger, and D.B.Pulling; Phys. Lett. {\bf A 207}, 109-112 (1995).

\bibitem{zou2}  Z. Zou, J. Ye, K. Oka, and Y. Nishihara, Phys. Rev. Lett.%
{\bf \ 80}, 1074 (1998).
\end{references}
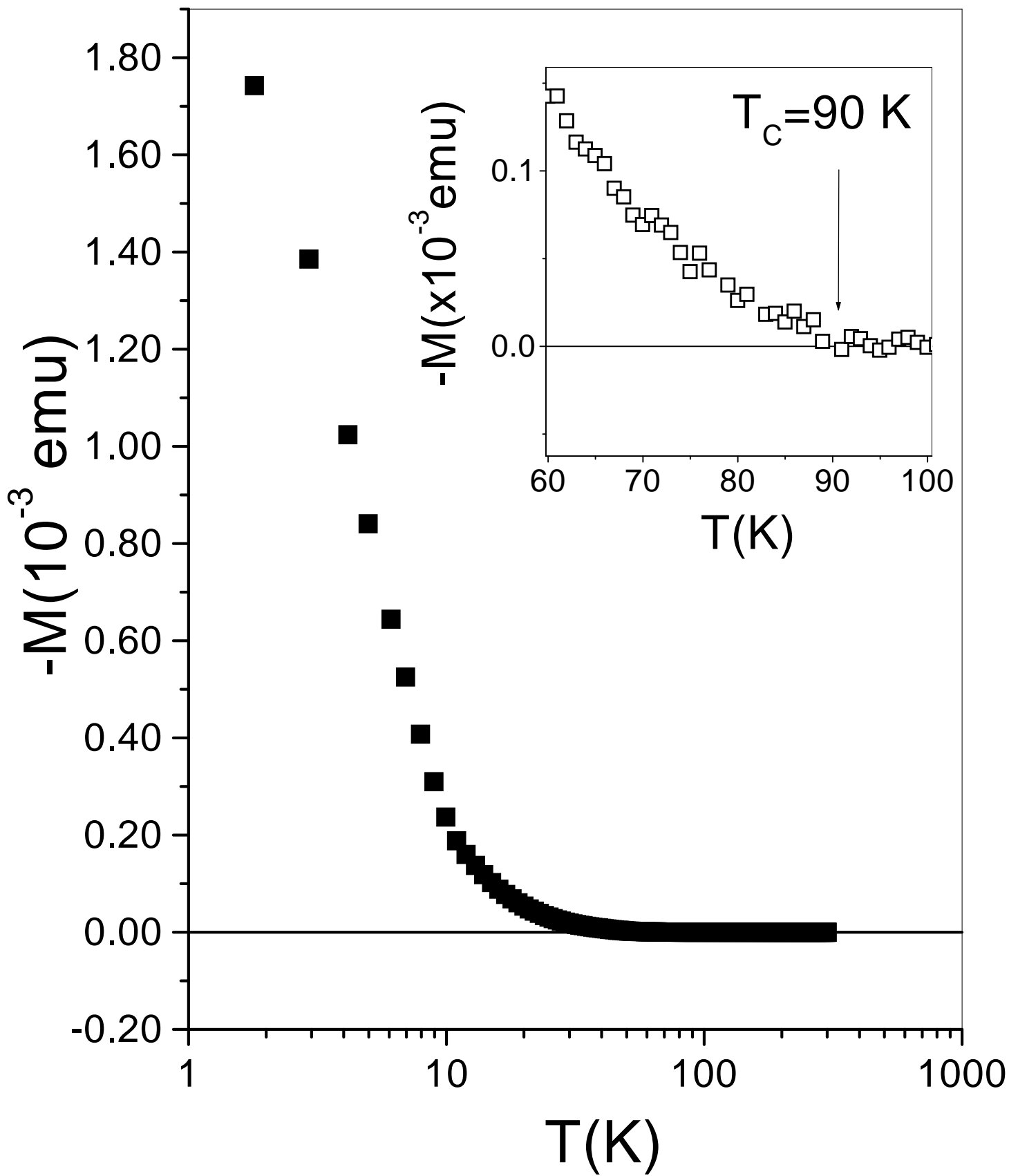
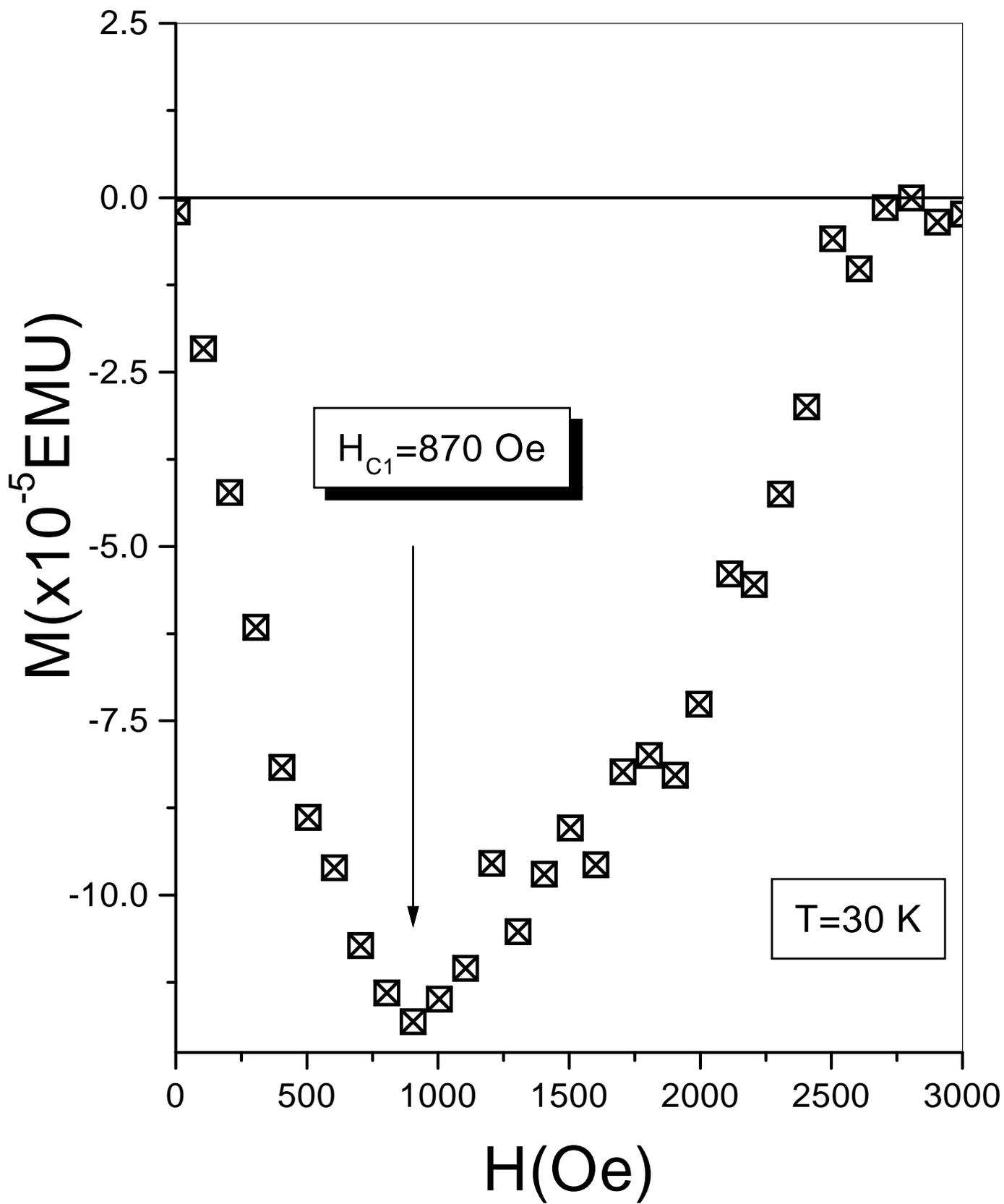
\end{document}